\documentclass{article}

    \usepackage[final]{neurips_2021}

\usepackage[utf8]{inputenc} 
\usepackage[T1]{fontenc}    
\usepackage{hyperref}       
\usepackage{url}            
\usepackage{booktabs}       
\usepackage{amsfonts}       
\usepackage{nicefrac}       
\usepackage{microtype}      
\usepackage{xcolor}         

\usepackage{graphicx}
\usepackage{physics}
\usepackage{array}
\usepackage{float}

\title{Randomized Benchmarking of Local Zeroth-Order Optimizers for Variational Quantum Systems}

\author{%
  Lucas Tecot \& Cho-Jui Hsieh \\
  Department of Computer Science\\
  University of California, Los Angeles\\
  Los Angeles, CA, USA \\
  \texttt{\{lucastecot, chohsieh\}@cs.ucla.edu} \\
}

\begin{document}

\maketitle

\begin{abstract}
    In the field of quantum information, classical optimizers play an important role. From experimentalists optimizing their physical devices to theorists exploring variational quantum algorithms, many aspects of quantum information require the use of a classical optimizer. For this reason, there are many papers that benchmark the effectiveness of different optimizers for specific quantum optimization tasks and choices of parameterized algorithms. However, for researchers exploring new algorithms or physical devices, the insights from these studies don't necessarily translate. To address this concern, we compare the performance of classical optimizers across a series of partially-randomized tasks to more broadly sample the space of quantum optimization problems. We focus on local zeroth-order optimizers due to their generally favorable performance and query-efficiency on quantum systems. We discuss insights from these experiments that can help motivate future works to improve these optimizers for use on quantum systems.
\end{abstract}

\section{Introduction}

Quantum computing has over time gathered more and more attention from researchers for the promise of significant computational speedups relative to classical computers. This has spurred many developments across all fronts in the field, from algorithms to building real quantum computers. However, many of these works still rely on the use of classical optimizers. For instance, variational quantum algorithms are a class of algorithms that have parameters that are then optimized by a classical optimizer \citep{cerezo_variational_2021}. These algorithms exist both as a way to do automated optimization of a quantum algorithm and to realize practically useful algorithms in noisy near-term devices.

And beyond the scope of algorithm design, classical optimizers also play a part in assisting experimentalists working in quantum information. Beyond simply being used in practice to realize theorized algorithms on real devices \citep{ebadi_quantum_2022}, optimizers can tune the control of physical actions (such as laser pulses, injections of electrical current, etc.) that all need to be controlled precisely to produce desired quantum operation \citep{coopmans_protocol_2021, leng_efficient_2023}. 

Furthermore, there are unique constraints imposed on optimizers of quantum systems that make this a unique problem. For instance, obtaining gradient information is in general much harder than in classical systems. Due to measurement collapse in quantum systems, in general it is impossible to achieve the same computational complexity as classical backpropagation. And it is only known how to achieve the same complexity if you have $O(polylog(M))$ copies of your input state, where $M$ is the number of parameters \citep{abbas_quantum_2023}. While there is still active work in this area, the difficulty of this problem makes zeroth-order methods that only sample and do not depend on gradients more appealing than they normally are in classical optimization problems.

For these reasons, understanding how classical optimizers interact with quantum objects is both difficult and highly important for designing the best quantum devices / algorithms. As a result, there are many studies on exactly that. Some of these studies are general benchmarks that compare a wide variety of optimizers \citep{pellow-jarman_comparison_2021, anand_natural_2021, singh_benchmarking_2023}; others are works that propose new optimizers for quantum circuits and show experimental evidence for the advantages of their optimizer \citep{sung_using_2020, gacon_simultaneous_2021, leng_efficient_2023}. However, most of these works benchmark for a fixed set of problems using a specific form of parameterized quantum ansatzes / models intended for each problem.

While these works are important for understanding specific use cases, it's not necessarily clear how much the insight from these works translates to new scenarios. For instance, if a researcher is developing a new variational quantum algorithm, how will they know which optimizer is likely to perform well? Even if the algorithm is familiar to one with existing optimizer benchmarks, how can they be certain that their changes didn't cause a significant distribution shift?
And in the case of physics experimentalists, similar concerns arise. Experimenalitsts can have varying devices where the type of control they have over their system are constrained in different ways. And there may be noise or aspects of their system that shift over time \citep{proctor_detecting_2020, blume-kohout_wildcard_2020}. So unless they constantly re-evaluate many optimizers on their own, what sort of confidence can they have in their choice of optimizer being wise?

These are the questions we take a shot at addressing in this work. We do this by benchmarking on tasks that are randomized. So in addition to the random parameter initialization common in other benchmarks, we also randomize the parameterized circuit / ansatz that is used, and in some cases randomize parts of the objective we are trying to minimize. While this is certainly not the most perfect way to answer the questions we posed before, as we have no way of knowing if our choice of randomization truly accurately benchmarks the distribution of useful optimization problems, we hope that by adding more variety via randomness in our benchmark we can begin to identify features of optimization algorithms that work more generically on many types of variational quantum problems.

However, when doing a study like this, you run into the risk of benchmarking something so generic that it's difficult to get any concrete insights from results. For this reason, we narrow our focus onto understanding how to improve a specific class of optimizers. First, we only consider \textbf{zeroth-order} methods. As mentioned previously, this is because methods that only sample and don't require gradients currently tend to be more easily realizable on quantum systems. Second, we use only \textbf{local} optimizers. This means that our optimizers sample the objective centered around a specific "canidate" point. Last, we only consider \textbf{sample-efficient} methods. This means that our optimizers make optimization decisions based on sampling as few points as possible per step.
These choices mostly centered around us deciding to focus on studying the SPSA algorithm \citep{spall_implementation_1998} and optimizers like it, for a handful of reasons. First, in general SPSA tends to perform competitively with most other optimizers in most existing benchmarks. Second, SPSA and methods like it are efficient to run on quantum systems and do not have runtime dependencies on the number of parameters you are optimizing. Third, while many zeroth-order methods have a hyperparameter for the number of sampled candidate parameters (and in the case of quantum, number of circuit evaluations) per optimization step, we desired to isolate this consideration for study in future works. We instead try to understand how we can push the limits of optimizers that aggressively take steps with minimal information-gain cost per step. (But we acknowledge that more deeply studying this aspect is highly important for future works.)

In this study, we benchmark randomized experiments for a variety of Hamiltonain minimzation and generative modeling tasks. We side-by-side compare the performance of 7 optimizers: SPSA, AdamSPSA, 2-SPSA, QNSPSA, GES, xNES, and sNES. We produce plots illustrating both the average rate of convergence and statistics on the end-result performance of each optimizer, and discuss our thoughts on insights to be gained from these results. But in short, we believe there are two \textbf{\textit{main take-aways}}. First, \textbf{more sophisticated optimizers are not generally better}. In our benchmarks SPSA tends to perform best overall, followed by the other simpler heuristic methods like AdamSPSA and GES. There is more nuance to this statement and it certainly isn't true in all circumstances, but under our randomized tasks methods simpler methods tended to be more reliably effective. Second, \textbf{there is a need for more robust or adaptive optimization heuristics}. While there are heurstics that can assist these optimizers in optimizing quickly in certain parts of the optimization process, at other points these heuristics can begin to hurt the optimizer's performance. As such, we argue that it would be beneficial to make these heuristics more robust to distribution shifts.
 But more broadly, we hope that this work helps stimulate thought into how to more broadly compare and study optimizers of quantum systems.

The code for this work can be found at \url{https://github.com/ltecot/rand_bench_opt_quantum}.

\section{Prior Work}

There are a number of works that explicitly benchmark a variety of optimizers. \citet{pellow-jarman_comparison_2021} compares a variety of both gradient and gradient-free optimizers on variational quantum linear solver problems, both in the presence and absense of noise. They show that while there's no clear best otpimizer, SPSA tends to perform favorably in realistic noise scenarios. \citet{anand_natural_2021} benchmarks natural evolutionary strategies  (NES) on variational quantum eigensolver (VQE) and state preparation problems. They also empirically investigate and provide some justification for how NES could be used in a hybrid algorithm to assist gradient-based optimizers in barren plateau regimes. \cite{singh_benchmarking_2023} benchmarks optimizers for a variety of quantum chemistry tasks. Like other studies, there's no clear best algorithm, but SPSA tends to perform well in noisy conditions.

Additionally, while not explicitly a benchmark, a number of works compare optimizers against a variety of tasks. \citep{sung_using_2020} introduces methods that use quadratic fitting of sampled points to evaluate the gradient and perform gradient descent and policy gradient descent. They additionally benchmark these methods against a variety of optimizers for three unique Hamiltonian-minimization problems with specific ansatzes. They also include some more practical considerations, such as the cost of evaluating different Hamiltonian measurments, the possibility of parallelizing multiple quantum circuit evaluations, and doing more robust hyperparameter tuning. In their results their method performs best, but SPSA can come close and often out-performs other methods in success rates. \citep{gacon_simultaneous_2021}, which proposes the QNSPSA algorithm we use in this study, also compares its performance to original SPSA on a variety of tasks and compares robustness to parameter initialization. \citep{leng_efficient_2023} does the same for their proposed AdamSPSA to SPSA and similar finite-difference methods, but they instead compare on the task of tuning the performance of a qubit operation on quantum computer.

\section{Benchmarks} \label{sec:benchmarks}

In this section we outline each of the benchmarks we perform in this paper, motivate the reasoning behind each experiment choice, and provide the finer details of each. While we overall strive to include some aspect of randomness / broadness, we make a few distinct choice of fixing specific elements between different benchmarks. Some of these choices are just so we can help distinguish any differences between types of problems, and others are so we try to understand differences between different levels of difficulty within a type of problem.

To ensure fairness, for all experimental runs, all randomness (initialization parameters + random circuits / Hamiltonians / distributions) are controlled by the random key. So although each run is randomly sampled, because we use the same random keys across all optimizers, they all run the same variations of each problem. The statistics of each run only vary as a result of the differences between each optimizer.

We also want to ensure that our results aren't biased by a poor choice of hyperparameters. However, especially because we are doing highly randomized tasks, it's difficult to identify what's a "good" choice of hyperparameter means. And even if they could be identified, it's not always reasonable to assume the user of said optimizer would be able to properly find them. Our compromise is to do hyperparameter tuning only on a small subset of the possible problem space. We select the hyperparameters according to a random search run on 3 random keys. This means that as we try random hyperparmeter combinations, they will be tried on 3 different random configurations of the optimization problem. So while this is not as exhaustive as the 100 we test on, the tuning isn't heavily biased to a single random problem sample. When there is a range of hyperparameters that all perform optimally, we bias our choice towards the default values typically used by the algorithm's authors or commonly selected in the literature. Once we have the tuned hyperparmeters, we benchmark each optimizer on each problem using 100 runs. All models in all experiments have their parameters initialized from a normal distribution of mean 0 and standard deviation $\pi$. 

\subsection{Hamiltonian Minimization Experiments} \label{sec:hamiltonian_min}

First we run experiments on Hamiltonain minimization problems. This means that we choose a Hamiltonian as an observable, and the expected value of measuring this Hamiltonian becomes the "loss" with which we are aiming to produce a quantum state that minimizes this loss. These benchmarks are meant to encapsulate use-cases such as variational quantum eigensolvers and quantum optimization problems that map some problem to a specific Hamiltonian. We produce our candidate states by parameterizing a quantum circuit and optimizing it to map a simple state (usually $\ket{0}$) to the state we measure with the Hamiltonian observable.

It is also important to note that we are simulating the noise-free version of this problem, as we assume we have access to the exact expected value of the Hamiltonian to minimize. While this is certainly not a realistic assumption, we wanted to first focus on how the aspects of each optimizer affect performance on quantum systems first before considering varying levels of noise as a factor.

All of the experiments we run here use random circuits as their quantum circuit / ansatz. Specifically we use the RandomLayers circuit by Pennylane \citep{bergholm_pennylane_2022}, which are layers of randomly placed parameterized single qubit X, Y, or Z rotation gates mixed with randomly placed CNOT gates.

\textbf{1D Ising Model: }
The first set of experiments we run use the 1D Ising model as our Hamiltonian observable. Specifically, we use
\begin{equation*}
    \mathcal{H} = - \sum_{i=1}^{N} Z_i \otimes Z_{i+1} - \frac{1}{2} \sum_{i=1}^{N} X_i.
\end{equation*}
As the 1D ising model is known to be an easily solvable problem, the intention of these experiments is to provide a simple baseline to understand how our optimizers perform on easier quantum optimization problems.
These experiments are run on systems of 3 qubits, with circuits of 30 parameterized single qubit gates and 10 2-qubit gates. This is relatively simpler and over-parameterized compared to the other experiments in this section. Each run is executed for 500 update steps of the optimizer.

\textbf{2D Heisenburg Model: }
Next we benchmark our optimizers on the 2D Heisenburg model. This serves as our harder problem, as not only do we include additional interaction terms w.r.t. the 1D Ising model, but now we also increase the dimensionality of the connectivity of our observable to a 2D lattice. We specifically use
\begin{align*}
    \mathcal{H} = - \frac{1}{2} \sum_{i=1}^{N} \sum_{j=1}^{N} \sum_{M \in \{X, Y, Z\}} M_{i,j} \otimes M_{i+1,j} + M_{i,j} \otimes M_{i,j+1} - \frac{1}{4} \sum_{i=1}^{N} \sum_{j=1}^{N} Z_{i,j}.
\end{align*}
In contrast to our 1D Ising experiments, these experiments are intended to gain insight into how our optimizers perform on a much more difficult problem. 
We use 9 qubit systems in these experiments, with quantum circuits containing 162 parameterized single qubit gates and 49 2-qubit gates. Each run is executed for 2000 update steps of the optimizer. One other import distinction is that these runs were hyper-parameter tuned for 1000 update steps, but we increased the experiments to 2000 steps because a few runs seemed to not be fully converging. We believe this lead to some interesting side-effects which we discuss in section \ref{sec:insights}.

\textbf{Randomized Hamiltonians: }
Lastly, we run experiments on randomized Hamiltonians. Specifically, we generate each Hamiltonian by combining single qubit measurment terms (randomly sampled Pauli X, Y, or Z gates on random qubits) with 2-qubit measurment terms (tensor products of randomly sampled Pauli X, Y, or Z gates on random qubits). We define the Hamiltonian as
\begin{gather*}
    \mathcal{H} = \sum_{i=1}^{N_d} c_i (A_i \otimes B_i) + \sum_{i=1}^{N_s} s_i S_i \\
    c_i, s_i \sim \mathcal{N}(0, \pi) \quad A_i, B_i, S_i \sim \mathcal{U}(\bigcup_{i \in [N]}\{X_i, Y_i, Z_i\}).
\end{gather*}
This benchmark exists for two purposes. First, it is intended to be a problem that is of "medium hardness" that is in-between the 1D Ising and the 2D Heisenburg experiments. Second, by adding randomness not only to the quantum circuit but also the objective, we hope to gain some additional coverage of many possible hamiltonian minimization problems than we did through the prior experiments to see if the insights from them have some evidence of generalization.
These experiments are run on systems of 10 qubits. For each run we randmomly sample 10 single qubit terms and 20 2-qubit terms to construct every Hamiltonian. Note that this means, unlike the prior two set of experiments, the energy objective for each optimization procedure differs between run to run. For the random circuit ansatz, we use 30 parameterized single qubit gates and 10 2-qubit gates. Each run is performed for 500 update steps.

\subsection{Quantum Generative Modelling Experiments}

To study a larger variety of variational quantum problems, we also investigate quantum generative modeling. In this setting, instead of trying to produce a state that minimizes and observable, we desire to produce a state that when measured in the full computational basis, matches a provided target distribution. Like with Hamiltonian minimization, we produce our candidate states by using a parameterized circuit. For the same reasons mentioned in section \ref{sec:hamiltonian_min}, we focus on the noise-less setting where we assume we have direct access to the true loss function. In this case we use the negative log-likelihood (NLL) loss.

\textbf{Cardinality Constrained Distribution + QCBM: }
Our first generative experiment is using the Quantum circuit Born machine \citep{benedetti_generative_2019} to generate a cardinality-constrained distribution. (So the only randomness in this benchmark is initalization parameters.) The purpose of this experiment is to serve as our baseline for quantum generative results. Because there aren't many optimizer benchmarks for quantum generative modelling, these results on a more standard test model and problem can help us interpret future more heavily randomized results. 
Specifically we run on a 10 qubit system, with 10 layers of 1 and 2 qubit gates of the QCBM ansatz (illustrated as $L=20$ in figure 1 of \citet{gili_quantum_2023}.) The cardinality we constrain to for our distribution is 5, meaning that our target distribution is the uniform distribution over any measurment of all 10 qubits that has 5 1's in the measurment result. Each run is executed for 5000 optimizer steps. 

\textbf{Randomized Distribution + Random Circuits: }
Our next generative experiment is a fully randomized problem. We use the same random layers ansatz used in section \ref{sec:hamiltonian_min}, and our target distribution is fully random. Specifically we use the absolute value of a normal distribution with mean 0 and standard deviation $\pi$, and then divide by the sum to normalize it to a valid probability distribution. In contrast to the other generative modelling experiment, this experiment exists to try to broadly sample many possible generative models and target distributions. 
For these experiments we run on 5 qubit systems, with 100 parameterized single qubit gates and 30 2-qubit gates in the random layers ansatz. Each run is executed for 5000 optimizer steps.

\section{Optimizers}

In this section we briefly outline all of the optimizers we benchmarked. While this is certainly not a fully exhaustive comparison of all local zeroth-order optimizers, we chose this selection because they cover most methods and heuristics used in SPSA-like methods. Additionally, most of these optimizers have a history of being used for parameterized quantum circuit tasks. 
Table \ref{tab:optimizers} in the appendix  shows the detailed update rules of these optimizers. 

\textbf{Simultaneous Pertubation Stochastic Approximation (SPSA)} \citep{spall_multivariate_1992}
 is a commonly used method, both inside and outside the context of optimizing quantum circuits. In a nutshell, SPSA is approximated gradient descent where we randomly sample directions in parameter space. Per step it samples a small random vector from a Rademacher distribution in parameter space, estimates the gradient along that vector with finite difference approximation, and then takes a step along said vector according to the sampled gradient to minimize loss. SPSA is also often used with learning rate and finite difference step size decay, which we also use here.

\textbf{AdamSPSA} \citep{leng_efficient_2023}
 is the application of the Adam optimizer heuristic \citep{kingma_adam_2017} on the SPSA algorithm. Specifically, it estimates via a running sum and updates according to momentum and variance normalization terms.

\textbf{2-SPSA} \citep{spall_accelerated_1997}
 is essentially an approximation of Newton's method, which is gradient descent where the gradient is multiplied by the inverse of the Hessian. To approximate the Hessian, it samples two random vectors (with a Rademacher distribution like in regular SPSA) and evaluates the 2nd order derivative along those two vectors. It then uses a weighted averaging of these samples to provide the Hessian used during optimization. Additionally, because the Hessian estimate can lead to more unstable updates, 2-SPSA also often blocks updates which increase the loss over a certain threshold from the prior value per step. 

\textbf{Quantum Natural SPSA (QNSPSA)} \citep{gacon_simultaneous_2021}
 is a variation of SPSA that utilizes the quantum natural gradient. Similar to classical natural information, using this metric has a few theoretical advantages to help accelerate and stabilize optimization. In practice, this method functions near-identically to 2-SPSA, except that when they sample the metric matrix, they compute the Hessian of the Fubini-Study metric instead of the Hessian of the loss function.

\textbf{Guided evolutionary strategies (GES)} \citep{maheswaranathan_guided_2019}
 is an evolutionary method with heuristic guiding. However, despite the different name, it is fundamentally very similar to SPSA with only two major differences. First, GES instead samples its random parameter-space vectors with a Gaussian instead of a Rademacher distribution. Second, GES biases the covariance of the sampling Gaussian along the subspace of the recent prior gradients. The intuition behind this choice is that, similar to momentum, that future gradients are more likely that not to be biased in the direction of the most recent prior gradients. However, instead of just increasing the update size in these directions, GES biases the sampling in this direction to increase information gain in this biased direction-of-travel.

\textbf{Exponential Natural Evolutionary Strategies (xNES)} \citep{wierstra_natural_2011}
 is an extension of evolutionary strategies to improve trainability. In this algorithm, it is assumed we have a multi-variate Gaussian in our space of model parameters, and our goal is to optimize this Gaussian to, in expectation, sample parameters that produce the lowest loss on the underlying problem. This is done by performing stochastic gradient descent on the parameters of the multi-variate Gaussian. xNES then augments this by instead using natural gradient descent to improve convergence guarantees, and utilizes an exponential matrix mapping to make the algorithm more computationally efficient.

\textbf{Seperable Natural Evolutionary Strategies (sNES)} \citep{wierstra_natural_2011} is simply xNES that assumes independence between parameters in order to be even more computationally efficient. It is functionally equivalent to xNES except where the covariance matrix only allows terms along the diagonal.

\section{Results}

\begin{figure}[h]
\begin{center}
\includegraphics[trim={7 0 7 0},clip,height=0.22\textheight]{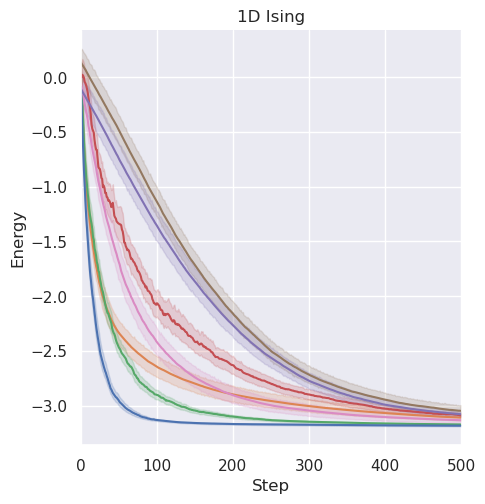}
\includegraphics[trim={19 0 7 0},clip,height=0.22\textheight]{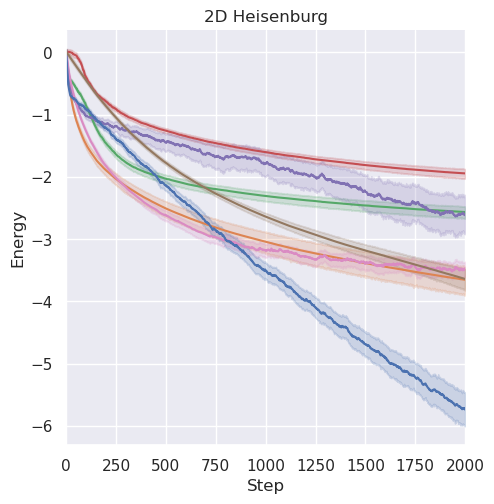}
\includegraphics[trim={19 0 7 0},clip,height=0.22\textheight]{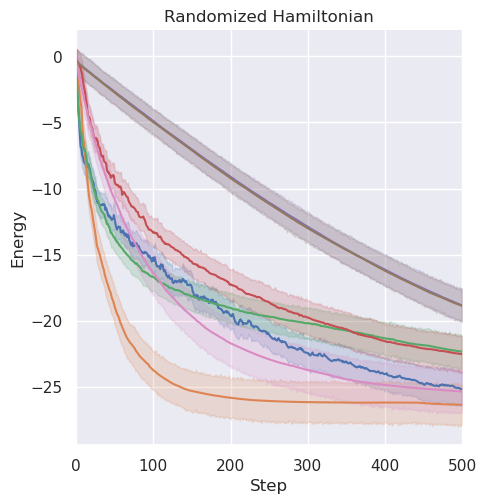}
\includegraphics[width=\textwidth]{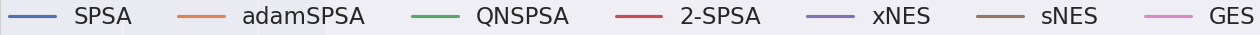}
\end{center}
\caption{Hamiltonian minimization experiment convergence plots w.r.t. number of optimizer steps. Plots mean value of all runs with the $95\%$ confidence interval. Experiment details can be found in section \ref{sec:benchmarks}.}
\label{plot:ham_converge}
\end{figure}

\begin{figure}[h]
\begin{center}
\includegraphics[trim={6 0 7 0},clip,height=0.24\textheight]{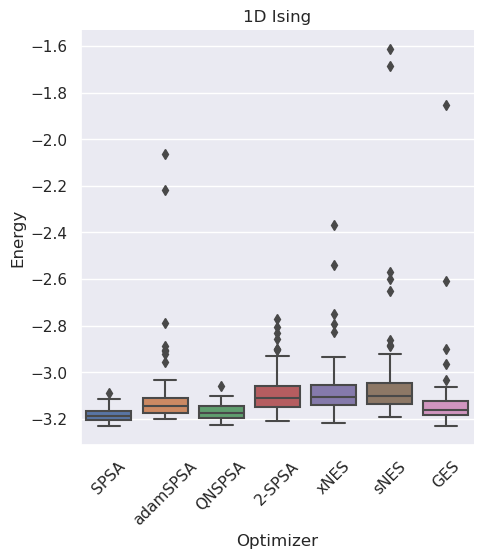}
\includegraphics[trim={19 0 7 0},clip,height=0.24\textheight]{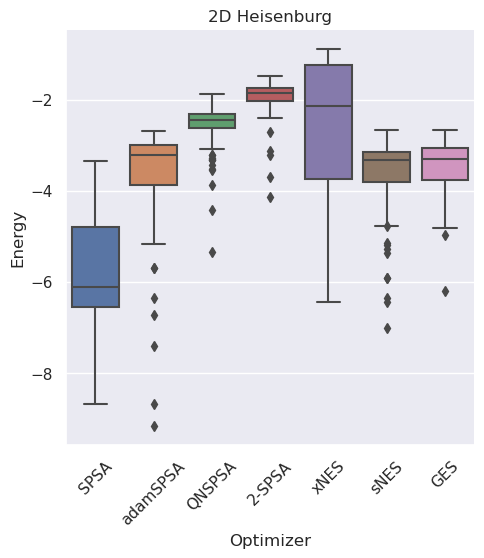}
\includegraphics[trim={19 0 7 0},clip,height=0.24\textheight]{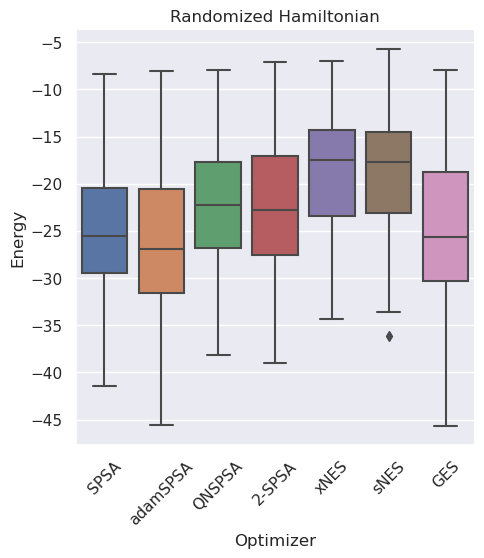}
\end{center}
\caption{Hamiltonian minimization experiment box plots. Plots the statistics of the final loss value from each run. Experiment details can be found in section \ref{sec:benchmarks}.}
\label{plot:ham_box}
\end{figure}

\begin{figure}[h]
\begin{center}
\includegraphics[trim={7 0 7 0},clip,height=0.28\textheight]{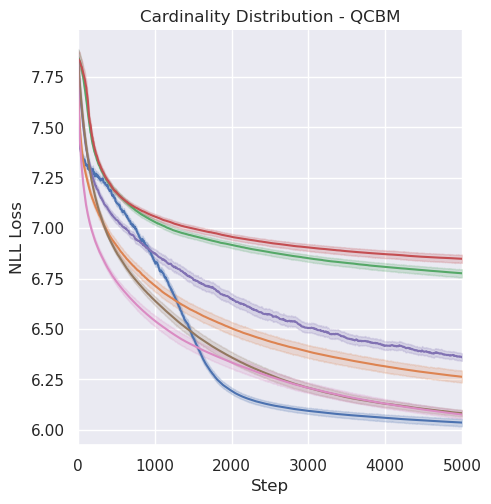}
\includegraphics[trim={19 0 7 0},clip,height=0.28\textheight]{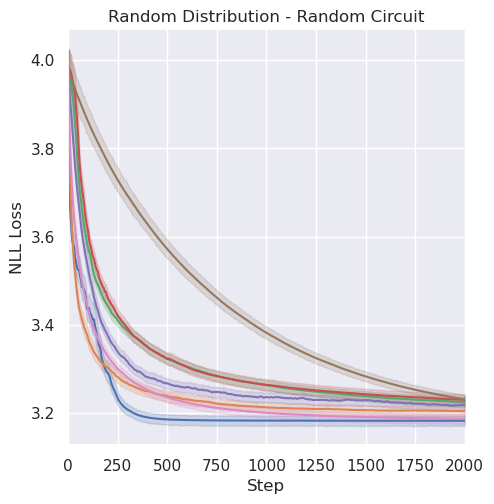}
\includegraphics[width=\textwidth]{figures/legend.png}
\end{center}
\caption{Generative modelling experiments convergence plots w.r.t. number of optimizer steps. Plots mean value of all runs with the $95\%$ confidence interval. Experiment details can be found in section \ref{sec:benchmarks}. Note that the NLL global minima is not always zero and can differ between problems.}
\label{plot:gen_converge}
\end{figure}

\begin{figure}[h]
\begin{center}
\includegraphics[trim={6 0 7 0},clip,height=0.31\textheight]{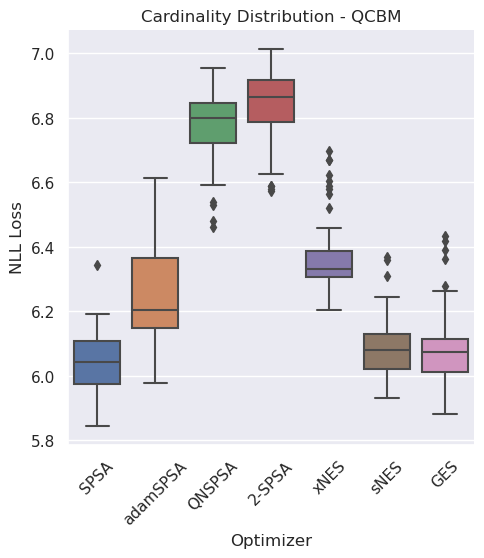}
\includegraphics[trim={19 0 7 0},clip,height=0.31\textheight]{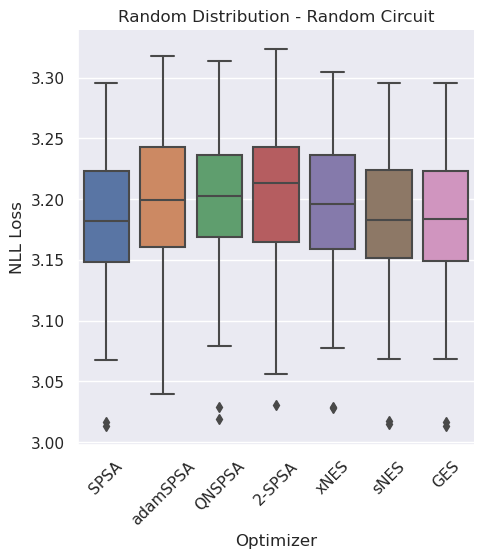}
\end{center}
\caption{Generative modelling experiments box plots. Plots the statistics of the final loss value from each run. Experiment details can be found in section \ref{sec:benchmarks}. Note that the NLL global minima is not always zero and can differ between problems.}
\label{plot:gen_box}
\end{figure}

We produce two types of plots for all experiments in section \ref{sec:benchmarks}: convergence plots and box plots. The convergence plots show the average loss of the optimization during each stage of the process, where the colored error area is the $95\%$ confidence interval of the mean. Note that while not every optimizer has the same amount of quantum computer query cost per step, how efficient each can be depends on specifics of the device and problem. So for the sake of this study we constrain each optimizer to sample as little as possible per step and assume this will result in only a small multiplicative factor difference in cost at each step.

The box plots illustrate the statistics of the end-result of each optimization run. In these plots, the center line of the colored region is the median loss value. The box region is the interquartile range (the range centered around the median that contains $50\%$ of the samples). The plot whiskers contain all points that are within $1.5$ times the size of the interquartile range from the median. All other points are considered outliers and are plotted individually.

The convergence plots are figures \ref{plot:ham_converge} and \ref{plot:gen_converge} for the Hamiltonian minimization and the generative modelling experiments respectively. The box plots are figures \ref{plot:ham_box} and \ref{plot:gen_box}, likewise for the hamiltonian minimization and the generative modelling experiments respectively. 

\subsection{Insights} \label{sec:insights}

There are a few main take-aways from these results that we believe these results illustrate. While we don't claim these results are concrete truths, we believe each of them warrant further study.

\textbf{1) Hyperparameter tuning and robustness for optimizer generalization is extremely important.} Generally this is illustrated by our results showing that most optimizer variants don't show clear benefits compared to their original versions when some hyperparameter tuning is done, which often contrasts the results shown in the original papers. However, this is more specifically illustrated in the convergence plot results of the 2D Heisenburg model shown in figure \ref{plot:ham_converge}. For this experiment, we hyperparameter tuned to optimizers taking 1000 steps but ran our experimental results out to 2000 steps. If you cut this plot off at 1000 steps, it would look very close to the random Hamiltonian experiments where SPSA, GES, and AdamSPSA perform similarly aside from the latter two converging more quickly. However, by choosing to optimize for longer SPSA is suddenly able to do significantly better than all optimizers. While this does raise the question of what a truly realistic hyperparameter tuning scenario is in quantum systems, we think it's likely more fruitful to sidestep this concern altogether and work to design optimizers that are adaptive or more robust to hyperparameter choice. 

\textbf{2) More complicated optimization strategies aren't generally better.} Despite that all of these optimizers can be viewed as additions to improve the general strategy of SPSA, in all of these benchmarks no optimizer clearly out-performs SPSA at the end of the optimization procedure. And the ones that come the closest are often ones like AdamSPSA and GES that rely on relatively simple and cheap heuristics. So while there is certainly something to be said about the theoretical benefits of using a method like QNSPSA, these results indicate that it may be possible to practically achieve better performance for cheaper by using simpler methods like step size decay and guiding heuristics.

\textbf{3) More complicated optimization strategies may be able to accelerate optimization.} It varies depending on the benchmark you look at, but often a few methods are able to converge to lower loss early on in the optimization procedure before being met or overtaken by SPSA. So while it may not be clearly better to use one of these methods as-is, these results indicate it might be possible to develop new adaptive methods to improve the convergence speed of methods like SPSA while not sacrificing overall performance.

\section{Outlook}

\subsection{Future Works} \label{sec:future_works}

Beyond the straightforward expansions on this work (larger studies, more optimizers, more relevant randomization, etc.) and works related to the insights in section \ref{sec:insights}, we feel the following would be particularly fruitful follow up studies:

\textbf{1) Noise -} This benchmark only compared the noiseless setting, but in practice noise is almost always involved with a quantum computer. Although there are papers that look at shot noise and how to select the right amount of circuit evaluations to still optimize efficiently \citep{gu_adaptive_2021}, we would like to see a study similar to the one we've done here that adds noise as an additional dimension to study where the transitions of an optimizer performing well v.s. poorly occurs.

\textbf{2) Sampling Amount -} For this study, we restricted ourselves to considering only sampling from a minimal number of parameter-difference vectors to get the information needed to take an update step. However, it's possible we could do better by relaxing this constraint. (It's especially surprising methods like GES worked as well as they did in this study, as they were intended to be used with multiple samples per step.) And on the other side of this, can we re-design some aspects of methods that typically benefit from additional samples per step to instead work well in a small-sample setting by aggregating information between steps? Both considerations would be critical for understanding the truly best strategies in this class of optimizer.

\textbf{3) Sampling Distribution -} The optimizers we study here use either a Gaussian or a Rademacher distribution to sample parameter vectors. While they both seem to be able to produce effective optimizers, studying in a more principled way the effects of different choices of probability distributions would be interesting. This could especially become more relevant if we look at optimizing parameters that add discrete constraints, as would be the case in some experimental setups, error-corrected quantum computation, coordinate descent parameter-shift rule based optimizers \citep{schuld_evaluating_2019}, or other more exotic forms of parameterization.

\textbf{4) Adaptive Methods - } Along the lines of what was mentioned in section \ref{sec:insights}, studying methods that adaptively change during the optimization process could be fruitful for a number of reasons. First, it could help combine the benefits of multiple strategies. Beyond the optimizers we covered here, there are works that find other ways to accelerate optimization. (For instance, \citet{luo_koopman_2022} use machine learning to predict optimization trajectories, and \citet{fontana_classical_2023}  classically models the loss landscape of certain parameterized circuits.) When it's possible to combine the information from quantum computer queries into multiple methods, having a strategy that can adapt to rely either more or less on a specific strategy during parts of the optimization process could allow us to have desirable properties of multiple methods (speed of convergence, ability to optimize well in difficult landscapes, flexibility of a method to work without prior assumptions, robustness to distribution shift, etc.) with a minimal cost-regret overhead. Second, such a method could provide insights into the limits of each of the above methods. By studying in which parts of an optimization process one method begins to be unable to optimize as well as another, it could provide insight to researchers looking to mathematically understand and characterize optimizers and loss landscapes. Lastly, such methods would make future benchmarking studies much simpler. Instead of having to be concerned about what reasonable hyperparameter tuning is and expending the resources to perform it, benchmarks could just compare adaptive versions of the methods in question. (And if said method has a regret bound, they can have precise confidence in the robustness of their results.)

\subsection{Conclusion}

In this work, we benchmark SPSA-like optimizers on a variety of parameterized quantum optimization tasks with randomized quantum circuits and randomized objectives. These results provide evidence to suggest that certain heuristics can help accelerate optimization, they often do not perform better than the simpler methods overall. However, we believe that in a broader sense this study helps illustrate the need not only for methods that are adaptive / more robust to hyperparameter choice, but also for broader thought on how we can effectively compare optimizers in quantum systems that aligns with the realistic scenarios they will be used in. While the take-aways from this work are intuitions and ultimately only serve to inform directions of future study, our hope is that this work will inspire more thought into how to best categorize and compare optimizers in variational quantum systems as a whole.


\subsubsection*{Acknowledgments}
The authors thank Kaitlin Gili, Ntwali Bashige, Chris Chien, and Di Luo for their insightful discussions and feedback.

\bibliography{references}

\begin{thebibliography}{25}
\providecommand{\natexlab}[1]{#1}
\providecommand{\url}[1]{\texttt{#1}}
\expandafter\ifx\csname urlstyle\endcsname\relax
  \providecommand{\doi}[1]{doi: #1}\else
  \providecommand{\doi}{doi: \begingroup \urlstyle{rm}\Url}\fi

\bibitem[Abbas et~al.(2023)Abbas, King, Huang, Huggins, Movassagh, Gilboa, and McClean]{abbas_quantum_2023}
Amira Abbas, Robbie King, Hsin-Yuan Huang, William~J. Huggins, Ramis Movassagh, Dar Gilboa, and Jarrod~R. McClean.
\newblock On quantum backpropagation, information reuse, and cheating measurement collapse, May 2023.
\newblock URL \url{http://arxiv.org/abs/2305.13362}.
\newblock arXiv:2305.13362 [quant-ph].

\bibitem[Anand et~al.(2021)Anand, Degroote, and Aspuru-Guzik]{anand_natural_2021}
Abhinav Anand, Matthias Degroote, and Alán Aspuru-Guzik.
\newblock Natural evolutionary strategies for variational quantum computation.
\newblock \emph{Machine Learning: Science and Technology}, 2\penalty0 (4):\penalty0 045012, July 2021.
\newblock ISSN 2632-2153.
\newblock \doi{10.1088/2632-2153/abf3ac}.
\newblock URL \url{https://dx.doi.org/10.1088/2632-2153/abf3ac}.
\newblock Publisher: IOP Publishing.

\bibitem[Benedetti et~al.(2019)Benedetti, Garcia-Pintos, Perdomo, Leyton-Ortega, Nam, and Perdomo-Ortiz]{benedetti_generative_2019}
Marcello Benedetti, Delfina Garcia-Pintos, Oscar Perdomo, Vicente Leyton-Ortega, Yunseong Nam, and Alejandro Perdomo-Ortiz.
\newblock A generative modeling approach for benchmarking and training shallow quantum circuits.
\newblock \emph{npj Quantum Information}, 5\penalty0 (1):\penalty0 1--9, May 2019.
\newblock ISSN 2056-6387.
\newblock \doi{10.1038/s41534-019-0157-8}.
\newblock URL \url{https://www.nature.com/articles/s41534-019-0157-8}.
\newblock Number: 1 Publisher: Nature Publishing Group.

\bibitem[Bergholm et~al.(2022)Bergholm, Izaac, Schuld, Gogolin, Ahmed, Ajith, Alam, Alonso-Linaje, AkashNarayanan, Asadi, Arrazola, Azad, Banning, Blank, Bromley, Cordier, Ceroni, Delgado, Di~Matteo, Dusko, Garg, Guala, Hayes, Hill, Ijaz, Isacsson, Ittah, Jahangiri, Jain, Jiang, Khandelwal, Kottmann, Lang, Lee, Loke, Lowe, McKiernan, Meyer, Montañez-Barrera, Moyard, Niu, O'Riordan, Oud, Panigrahi, Park, Polatajko, Quesada, Roberts, Sá, Schoch, Shi, Shu, Sim, Singh, Strandberg, Soni, Száva, Thabet, Vargas-Hernández, Vincent, Vitucci, Weber, Wierichs, Wiersema, Willmann, Wong, Zhang, and Killoran]{bergholm_pennylane_2022}
Ville Bergholm, Josh Izaac, Maria Schuld, Christian Gogolin, Shahnawaz Ahmed, Vishnu Ajith, M.~Sohaib Alam, Guillermo Alonso-Linaje, B.~AkashNarayanan, Ali Asadi, Juan~Miguel Arrazola, Utkarsh Azad, Sam Banning, Carsten Blank, Thomas~R. Bromley, Benjamin~A. Cordier, Jack Ceroni, Alain Delgado, Olivia Di~Matteo, Amintor Dusko, Tanya Garg, Diego Guala, Anthony Hayes, Ryan Hill, Aroosa Ijaz, Theodor Isacsson, David Ittah, Soran Jahangiri, Prateek Jain, Edward Jiang, Ankit Khandelwal, Korbinian Kottmann, Robert~A. Lang, Christina Lee, Thomas Loke, Angus Lowe, Keri McKiernan, Johannes~Jakob Meyer, J.~A. Montañez-Barrera, Romain Moyard, Zeyue Niu, Lee~James O'Riordan, Steven Oud, Ashish Panigrahi, Chae-Yeun Park, Daniel Polatajko, Nicolás Quesada, Chase Roberts, Nahum Sá, Isidor Schoch, Borun Shi, Shuli Shu, Sukin Sim, Arshpreet Singh, Ingrid Strandberg, Jay Soni, Antal Száva, Slimane Thabet, Rodrigo~A. Vargas-Hernández, Trevor Vincent, Nicola Vitucci, Maurice Weber, David Wierichs, Roeland Wiersema, Moritz
  Willmann, Vincent Wong, Shaoming Zhang, and Nathan Killoran.
\newblock {PennyLane}: {Automatic} differentiation of hybrid quantum-classical computations, July 2022.
\newblock URL \url{http://arxiv.org/abs/1811.04968}.
\newblock arXiv:1811.04968 [physics, physics:quant-ph].

\bibitem[Blume-Kohout et~al.(2020)Blume-Kohout, Rudinger, Nielsen, Proctor, and Young]{blume-kohout_wildcard_2020}
Robin Blume-Kohout, Kenneth Rudinger, Erik Nielsen, Timothy Proctor, and Kevin Young.
\newblock Wildcard error: {Quantifying} unmodeled errors in quantum processors, December 2020.
\newblock URL \url{http://arxiv.org/abs/2012.12231}.
\newblock arXiv:2012.12231 [quant-ph].

\bibitem[Cerezo et~al.(2021)Cerezo, Arrasmith, Babbush, Benjamin, Endo, Fujii, McClean, Mitarai, Yuan, Cincio, and Coles]{cerezo_variational_2021}
M.~Cerezo, Andrew Arrasmith, Ryan Babbush, Simon~C. Benjamin, Suguru Endo, Keisuke Fujii, Jarrod~R. McClean, Kosuke Mitarai, Xiao Yuan, Lukasz Cincio, and Patrick~J. Coles.
\newblock Variational {Quantum} {Algorithms}.
\newblock \emph{Nature Reviews Physics}, 3\penalty0 (9):\penalty0 625--644, August 2021.
\newblock ISSN 2522-5820.
\newblock \doi{10.1038/s42254-021-00348-9}.
\newblock URL \url{http://arxiv.org/abs/2012.09265}.
\newblock arXiv:2012.09265 [quant-ph, stat].

\bibitem[Coopmans et~al.(2021)Coopmans, Luo, Kells, Clark, and Carrasquilla]{coopmans_protocol_2021}
Luuk Coopmans, Di~Luo, Graham Kells, Bryan~K. Clark, and Juan Carrasquilla.
\newblock Protocol {Discovery} for the {Quantum} {Control} of {Majoranas} by {Differentiable} {Programming} and {Natural} {Evolution} {Strategies}.
\newblock \emph{PRX Quantum}, 2\penalty0 (2):\penalty0 020332, June 2021.
\newblock ISSN 2691-3399.
\newblock \doi{10.1103/PRXQuantum.2.020332}.
\newblock URL \url{http://arxiv.org/abs/2008.09128}.
\newblock arXiv:2008.09128 [cond-mat, physics:physics, physics:quant-ph].

\bibitem[Ebadi et~al.(2022)Ebadi, Keesling, Cain, Wang, Levine, Bluvstein, Semeghini, Omran, Liu, Samajdar, Luo, Nash, Gao, Barak, Farhi, Sachdev, Gemelke, Zhou, Choi, Pichler, Wang, Greiner, Vuletic, and Lukin]{ebadi_quantum_2022}
Sepehr Ebadi, Alexander Keesling, Madelyn Cain, Tout~T. Wang, Harry Levine, Dolev Bluvstein, Giulia Semeghini, Ahmed Omran, Jinguo Liu, Rhine Samajdar, Xiu-Zhe Luo, Beatrice Nash, Xun Gao, Boaz Barak, Edward Farhi, Subir Sachdev, Nathan Gemelke, Leo Zhou, Soonwon Choi, Hannes Pichler, Shengtao Wang, Markus Greiner, Vladan Vuletic, and Mikhail~D. Lukin.
\newblock Quantum {Optimization} of {Maximum} {Independent} {Set} using {Rydberg} {Atom} {Arrays}.
\newblock \emph{Science}, 376\penalty0 (6598):\penalty0 1209--1215, June 2022.
\newblock ISSN 0036-8075, 1095-9203.
\newblock \doi{10.1126/science.abo6587}.
\newblock URL \url{http://arxiv.org/abs/2202.09372}.
\newblock arXiv:2202.09372 [cond-mat, physics:physics, physics:quant-ph].

\bibitem[Fontana et~al.(2023)Fontana, Rudolph, Duncan, Rungger, and Cîrstoiu]{fontana_classical_2023}
Enrico Fontana, Manuel~S. Rudolph, Ross Duncan, Ivan Rungger, and Cristina Cîrstoiu.
\newblock Classical simulations of noisy variational quantum circuits, June 2023.
\newblock URL \url{http://arxiv.org/abs/2306.05400}.
\newblock arXiv:2306.05400 [quant-ph].

\bibitem[Gacon et~al.(2021)Gacon, Zoufal, Carleo, and Woerner]{gacon_simultaneous_2021}
Julien Gacon, Christa Zoufal, Giuseppe Carleo, and Stefan Woerner.
\newblock Simultaneous {Perturbation} {Stochastic} {Approximation} of the {Quantum} {Fisher} {Information}.
\newblock \emph{Quantum}, 5:\penalty0 567, October 2021.
\newblock ISSN 2521-327X.
\newblock \doi{10.22331/q-2021-10-20-567}.
\newblock URL \url{http://arxiv.org/abs/2103.09232}.
\newblock arXiv:2103.09232 [quant-ph].

\bibitem[Gili et~al.(2023)Gili, Hibat-Allah, Mauri, Ballance, and Perdomo-Ortiz]{gili_quantum_2023}
Kaitlin Gili, Mohamed Hibat-Allah, Marta Mauri, Chris Ballance, and Alejandro Perdomo-Ortiz.
\newblock Do quantum circuit {Born} machines generalize?
\newblock \emph{Quantum Science and Technology}, 8\penalty0 (3):\penalty0 035021, May 2023.
\newblock ISSN 2058-9565.
\newblock \doi{10.1088/2058-9565/acd578}.
\newblock URL \url{https://dx.doi.org/10.1088/2058-9565/acd578}.
\newblock Publisher: IOP Publishing.

\bibitem[Gu et~al.(2021)Gu, Lowe, Dub, Coles, and Arrasmith]{gu_adaptive_2021}
Andi Gu, Angus Lowe, Pavel~A. Dub, Patrick~J. Coles, and Andrew Arrasmith.
\newblock Adaptive shot allocation for fast convergence in variational quantum algorithms.
\newblock 2021.
\newblock \doi{10.48550/arXiv.2108.10434}.
\newblock URL \url{http://arxiv.org/abs/2108.10434}.
\newblock arXiv:2108.10434 [quant-ph].

\bibitem[Kingma and Ba(2017)]{kingma_adam_2017}
Diederik~P. Kingma and Jimmy Ba.
\newblock Adam: {A} {Method} for {Stochastic} {Optimization}, January 2017.
\newblock URL \url{http://arxiv.org/abs/1412.6980}.
\newblock arXiv:1412.6980 [cs].

\bibitem[Leng et~al.(2023)Leng, Mundada, Ghadimi, and Houck]{leng_efficient_2023}
Zhaoqi Leng, Pranav Mundada, Saeed Ghadimi, and Andrew Houck.
\newblock Efficient {Algorithms} for {High}-{Dimensional} {Quantum} {Optimal} {Control} of a {Transmon} {Qubit}.
\newblock \emph{Physical Review Applied}, 19\penalty0 (4):\penalty0 044034, April 2023.
\newblock \doi{10.1103/PhysRevApplied.19.044034}.
\newblock URL \url{https://link.aps.org/doi/10.1103/PhysRevApplied.19.044034}.
\newblock Publisher: American Physical Society.

\bibitem[Luo et~al.(2022)Luo, Shen, Dangovski, and Soljačić]{luo_koopman_2022}
Di~Luo, Jiayu Shen, Rumen Dangovski, and Marin Soljačić.
\newblock Koopman {Operator} learning for {Accelerating} {Quantum} {Optimization} and {Machine} {Learning}, November 2022.
\newblock URL \url{https://arxiv.org/abs/2211.01365v1}.

\bibitem[Maheswaranathan et~al.(2019)Maheswaranathan, Metz, Tucker, Choi, and Sohl-Dickstein]{maheswaranathan_guided_2019}
Niru Maheswaranathan, Luke Metz, George Tucker, Dami Choi, and Jascha Sohl-Dickstein.
\newblock Guided evolutionary strategies: {Augmenting} random search with surrogate gradients, June 2019.
\newblock URL \url{http://arxiv.org/abs/1806.10230}.
\newblock arXiv:1806.10230 [cs, stat].

\bibitem[Pellow-Jarman et~al.(2021)Pellow-Jarman, Sinayskiy, Pillay, and Petruccione]{pellow-jarman_comparison_2021}
Aidan Pellow-Jarman, Ilya Sinayskiy, Anban Pillay, and Francesco Petruccione.
\newblock A comparison of various classical optimizers for a variational quantum linear solver.
\newblock \emph{Quantum Information Processing}, 20\penalty0 (6):\penalty0 202, June 2021.
\newblock ISSN 1573-1332.
\newblock \doi{10.1007/s11128-021-03140-x}.
\newblock URL \url{https://doi.org/10.1007/s11128-021-03140-x}.

\bibitem[Proctor et~al.(2020)Proctor, Revelle, Nielsen, Rudinger, Lobser, Maunz, Blume-Kohout, and Young]{proctor_detecting_2020}
Timothy Proctor, Melissa Revelle, Erik Nielsen, Kenneth Rudinger, Daniel Lobser, Peter Maunz, Robin Blume-Kohout, and Kevin Young.
\newblock Detecting and tracking drift in quantum information processors.
\newblock \emph{Nature Communications}, 11\penalty0 (1):\penalty0 5396, October 2020.
\newblock ISSN 2041-1723.
\newblock \doi{10.1038/s41467-020-19074-4}.
\newblock URL \url{https://www.nature.com/articles/s41467-020-19074-4}.

\bibitem[Schuld et~al.(2019)Schuld, Bergholm, Gogolin, Izaac, and Killoran]{schuld_evaluating_2019}
Maria Schuld, Ville Bergholm, Christian Gogolin, Josh Izaac, and Nathan Killoran.
\newblock Evaluating analytic gradients on quantum hardware.
\newblock \emph{Physical Review A}, 99\penalty0 (3):\penalty0 032331, March 2019.
\newblock ISSN 2469-9926, 2469-9934.
\newblock \doi{10.1103/PhysRevA.99.032331}.
\newblock URL \url{http://arxiv.org/abs/1811.11184}.
\newblock arXiv:1811.11184 [quant-ph].

\bibitem[Singh et~al.(2023)Singh, Mishra, and Majumder]{singh_benchmarking_2023}
Harshdeep Singh, Sabyashachi Mishra, and Sonjoy Majumder.
\newblock Benchmarking of {Different} {Optimizers} in the {Variational} {Quantum} {Algorithms} for {Applications} in {Quantum} {Chemistry}, February 2023.
\newblock URL \url{http://arxiv.org/abs/2208.10285}.
\newblock arXiv:2208.10285 [quant-ph].

\bibitem[Spall(1992)]{spall_multivariate_1992}
J.C. Spall.
\newblock Multivariate stochastic approximation using a simultaneous perturbation gradient approximation.
\newblock \emph{IEEE Transactions on Automatic Control}, 37\penalty0 (3):\penalty0 332--341, March 1992.
\newblock ISSN 1558-2523.
\newblock \doi{10.1109/9.119632}.
\newblock Conference Name: IEEE Transactions on Automatic Control.

\bibitem[Spall(1997)]{spall_accelerated_1997}
J.C. Spall.
\newblock Accelerated second-order stochastic optimization using only function measurements.
\newblock In \emph{Proceedings of the 36th {IEEE} {Conference} on {Decision} and {Control}}, volume~2, pages 1417--1424 vol.2, December 1997.
\newblock \doi{10.1109/CDC.1997.657661}.
\newblock ISSN: 0191-2216.

\bibitem[Spall(1998)]{spall_implementation_1998}
J.C. Spall.
\newblock Implementation of the simultaneous perturbation algorithm for stochastic optimization.
\newblock \emph{IEEE Transactions on Aerospace and Electronic Systems}, 34\penalty0 (3):\penalty0 817--823, July 1998.
\newblock ISSN 1557-9603.
\newblock \doi{10.1109/7.705889}.
\newblock Conference Name: IEEE Transactions on Aerospace and Electronic Systems.

\bibitem[Sung et~al.(2020)Sung, Yao, Harrigan, Rubin, Jiang, Lin, Babbush, and McClean]{sung_using_2020}
Kevin~J. Sung, Jiahao Yao, Matthew~P. Harrigan, Nicholas~C. Rubin, Zhang Jiang, Lin Lin, Ryan Babbush, and Jarrod~R. McClean.
\newblock Using models to improve optimizers for variational quantum algorithms.
\newblock \emph{Quantum Science and Technology}, 5\penalty0 (4):\penalty0 044008, October 2020.
\newblock ISSN 2058-9565.
\newblock \doi{10.1088/2058-9565/abb6d9}.
\newblock URL \url{http://arxiv.org/abs/2005.11011}.
\newblock arXiv:2005.11011 [quant-ph].

\bibitem[Wierstra et~al.(2011)Wierstra, Schaul, Glasmachers, Sun, and Schmidhuber]{wierstra_natural_2011}
Daan Wierstra, Tom Schaul, Tobias Glasmachers, Yi~Sun, and Jürgen Schmidhuber.
\newblock Natural {Evolution} {Strategies}, June 2011.
\newblock URL \url{http://arxiv.org/abs/1106.4487}.
\newblock arXiv:1106.4487 [cs, stat].

\end{thebibliography}
\bibliographystyle{plainnat}

\newpage

\appendix



\section*{Appendix}

\begin{table}[H]
    \centering
    \resizebox{!}{0.7\textwidth}{
    \begin{tabular}{ m{0.44\textwidth} m{0.44\textwidth} } 
      \begin{tabular}{c}
        \textbf{AdamSPSA} \\
        \hline \rule{0pt}{3ex} 
        \parbox{0.44\textwidth}{\begin{align*}
            &\Delta_i \sim \mathcal{U}(\{-1,1\}^d) \\
            & \epsilon_i = \epsilon_0 / i^{\gamma}, \eta_i = \eta_0 / (c + i)^{\alpha}, \beta_i = \beta_0 / i^{\lambda} \\
            &\nabla_{\text{est}} f(\theta_i) = \frac{f(\theta_i + \epsilon_i \Delta_i) - f(\theta_i - \epsilon_i \Delta_i)}{2 \epsilon_i} \\
            & \left(m_1 = \nabla_{\text{est}} f(\theta_1), v_1 = (\nabla_{\text{est}} f(\theta_1))^2\right)\\
            &m_i = \beta_i m_{i-1} + (1 - \beta_i) \nabla_{\text{est}} f(\theta_i) \\
            &v_i = \gamma v_{i-1} + (1 - \gamma) (\nabla_{\text{est}} f(\theta_i))^2 \\
            &\theta_{i+1} = \theta_i - \frac{\eta_i}{\sqrt{v_i} + \delta} m_i
        \end{align*}}
        \\
        \textbf{2-SPSA} \\
        \hline \rule{0pt}{3ex} 
        \parbox{0.44\textwidth}{\begin{align*}
            &\Delta_i, \Delta'_i \sim \mathcal{U}(\{-1,1\}^d) \\
            &\nabla_{\text{est}} f(\theta_i) = \frac{f(\theta_i + \epsilon_i \Delta_i) - f(\theta_i - \epsilon_i \Delta_i)}{2 \epsilon_i} \\
            &\delta f = f(\theta_i + \epsilon \Delta_i + \epsilon \Delta'_i) - f(\theta_i + \epsilon \Delta_i) \\
            & \qquad - f(\theta_i - \epsilon \Delta_i + \epsilon \Delta'_i) + f(\theta_i - \epsilon \Delta_i) \\
            &\hat{H}_i = \frac{\delta f}{2\epsilon^2} \frac{\Delta_i (\Delta'_i)^T + \Delta'_i (\Delta_i)^T}{2} \\
            &H_i = \frac{i}{i+1} H_{i-1} + \frac{1}{i+1} \hat{H}_i \\
            &\theta_{i+1} = \theta_i - \eta H_i^{-1} \nabla_{\text{est}} f(\theta_i)
        \end{align*}}
        \\
        \textbf{QNSPSA} \\
        \hline \rule{0pt}{3ex} 
        \parbox{0.44\textwidth}{\begin{align*}
            &\Delta_i, \Delta'_i \sim \mathcal{U}(\{-1,1\}^d) \\
            &F(\theta, \theta') = | \braket{\psi(\theta)}{\psi(\theta')} |^2 \\
            &\nabla_{\text{est}} f(\theta_i) = \frac{f(\theta_i + \epsilon_i \Delta_i) - f(\theta_i - \epsilon_i \Delta_i)}{2 \epsilon_i} \\
            &\delta F = F(\theta_i, \theta_i + \epsilon \Delta_i + \epsilon \Delta'_i) \\
            & \qquad - F(\theta_i, \theta_i + \epsilon \Delta_i) \\
            & \qquad - F(\theta_i, \theta_i - \epsilon \Delta_i + \epsilon \Delta'_i) \\
            & \qquad + F(\theta_i, \theta_i - \epsilon \Delta_i) \\
            &\hat{H}_i = \frac{-\delta F}{4\epsilon^2} \frac{\Delta_i (\Delta'_i)^T + \Delta'_i (\Delta_i)^T}{2} \\
            &H_i = \frac{i}{i+1} H_{i-1} + \frac{1}{i+1} \hat{H}_i \\
            &\theta_{i+1} = \theta_i - \eta H_i^{-1} \nabla_{\text{est}} f(\theta_i)
        \end{align*}}
      \end{tabular}
      & 
      \begin{tabular}{c}
        \textbf{SPSA} \\
        \hline \rule{0pt}{3ex} 
        \parbox{0.44\textwidth}{\begin{align*}
            &\Delta_i \sim \mathcal{U}(\{-1,1\}^d) \\
            & \epsilon_i = \epsilon_0 / i^{\gamma}, \eta_i = \eta_0 / (c + i)^{\alpha} \\
            &\nabla_{\text{est}} f(\theta_i) = \frac{f(\theta_i + \epsilon_i \Delta_i) - f(\theta_i - \epsilon_i \Delta_i)}{2 \epsilon_i} \\
            &\theta_{i+1} = \theta_i - \eta_i \nabla_{\text{est}} f(\theta_i)
        \end{align*}}
        \\
        \textbf{GES} \\
        \hline \rule{0pt}{3ex} 
        \parbox{0.44\textwidth}{\begin{align*}
            & U_i = \text{orthonormal basis of span of } \\
            & \qquad \{\nabla_{\text{est}} f(\theta_{i-k}), \dots, \nabla_{\text{est}} f(\theta_i)\}\\
            &\Sigma_i = \frac{\alpha}{n} I + \frac{1 - \alpha}{k} U U^T, \Sigma_{0,\dots,k} = \frac{1}{n} I \\
            &\Delta_i \sim \mathcal{N}(0, \sigma^2 \Sigma_i) \\
            &\nabla_{\text{est}} f(\theta_i) = \beta \frac{f(\theta_i + \Delta_i) - f(\theta_i - \Delta_i)}{2 \sigma^2} \\
            &\theta_{i+1} = \theta_i - \eta \nabla_{\text{est}} f(\theta_i)
        \end{align*}}
        \\
        \textbf{xNES} \\
        \hline \rule{0pt}{3ex} 
        \parbox{0.44\textwidth}{\begin{align*}
            &\Delta_i, \Delta'_i \sim \mathcal{N}(0, I), B_1 = I \\
            & z = \theta_i + \sigma_i B_i^T \Delta_i, z' = \theta_i + \sigma_i B_i^T \Delta'_i \\
            & u = 0.5 \text{ if } f(z) < f(z') \text{ else } -0.5 \\
            & u' = 0.5 \text{ if } f(z') < f(z) \text{ else } -0.5 \\
            &\nabla_{\mu} J = us + u's' \\
            &\nabla_{M} J = u(ss^T - I) + u'(s's'^T - I) \\
            &\nabla_{\sigma} J = \text{tr}(\nabla_{M} J) / d \\
            &\nabla_{B} J  = \nabla_{M} J - \nabla_{\sigma} J \cdot I \\
            &\sigma_{i+1} = \sigma_{i} \exp(\eta_{\sigma} / 2 \cdot \nabla_{\sigma} J) \\
            &B_{i+1} = B_{i} \exp(\eta_{B} / 2 \cdot \nabla_{B} J) \\
            &\theta_{i+1} = \theta_i + \eta_{\mu} \sigma_i B_i \nabla_{\mu} J \\
        \end{align*}}
        \\
        \textbf{sNES} \\
        \hline \rule{0pt}{3ex} 
        \parbox{0.44\textwidth}{\begin{align*}
            &\Delta_i, \Delta'_i \sim \mathcal{N}(0, I) \\
            & z = \theta_i + \sigma_i \Delta_i, z' = \theta_i + \sigma_i \Delta'_i \\
            & u = 0.5 \text{ if } f(z) < f(z') \text{ else } -0.5 \\
            & u' = 0.5 \text{ if } f(z') < f(z) \text{ else } -0.5 \\
            &\nabla_{\mu} J = us + u's' \\
            &\nabla_{\sigma} J = u(s^2 - 1) + u'(s'^2 - 1) \\
            &\sigma_{i+1} = \sigma_{i} \exp(\eta_{\sigma} / 2 \cdot \nabla_{\sigma} J) \\
            &\theta_{i+1} = \theta_i + \eta_{\mu} \sigma_i B_i \nabla_{\mu} J \\
        \end{align*}}
      \end{tabular} \\
    \end{tabular}
    }
    \caption{Optimizer algorithms. Illustrates a step of each optimizer in equation form. Note that these algorithms may slightly differ from the original works due to simplifying choices / constraints we made in this study. $f$ is the loss function and $\theta$ are the parameters per step. $\psi$ is the parameterized quantum circuit model used by $f$. $\mathcal{U}$ is the uniform distribution and $\mathcal{N}$ is the Gaussian distribution. $(\cdot)^2,\sqrt{(\cdot)}$ are element-wise on vectors. All other un-defined variables are hyperparameters.}
    \label{tab:optimizers}
\end{table}




\end{document}